\documentclass[]{spie}  

\usepackage{amsmath,amsfonts,amssymb}
\usepackage{graphicx}
\usepackage[colorlinks=true, allcolors=blue]{hyperref}

\title{Discovery space and science with the PLACID stellar coronagraph}

\author[a]{Ruben Tandon}
\author[a]{Liurong Lin}
\author[a]{Axel Potier}
\author[b]{Laurent Jolissaint}
\author[b]{Audrey Baur}
\author[c]{Derya \"Ozt\"urk Çetni}
\author[a]{Jonas G. K\"uhn}
\affil[a]{Department of Space and Planetary Sciences, University of Bern, Sidlerstrasse 5, 3012 Bern, Switzerland}
\affil[b]{University of Applied Sciences HEIG-VD, Route de Cheseaux 1, 1401 Yverdon-les-Bains, Switzerland}
\affil[c]{Atatürk University Astronomical Research and Application Center (ATASAM), Lütfü Ülkümen Street. No : 8/3, 25050, Yakutiye / Erzurum, Turkey}

\authorinfo{Further author information: \\ Ruben Tandon: E-mail: ruben.tandon@unibe.ch, Telephone: +41 31 684 32 90}

\begin{document} 
\maketitle

\begin{abstract}
The world’s first ever “adaptive stellar coronagraph” facility will be the PLACID instrument, installed on Turkey's new national observatory 4-m DAG telescope. PLACID incorporates a customized spatial light modulator (SLM) acting as a dynamically addressed focal-plane phase mask (FPM) coronagraph in the H – Ks bands. This new approach to high-contrast imaging will be applied on-sky in late 2024/early 2025. 
We present a first estimate of the science discovery space for PLACID, in terms of known exoplanets and brown dwarfs, considering raw lab contrast, contrast ratios, limiting magnitudes, coronagraphic inner working angle etc. In the future, we will also look into predicted disk and binary or multiple stars systems imaging performance, with the latter being a possible niche science case for the instrument (adaptive FPM for multiple stars). This work will inform on the first light PLACID commissioning activities and early science on the DAG telescope and is deemed to evolve in function of future developments on the DAG AO instrumentation suite. 
\end{abstract}

\keywords{Direct imaging, high-contrast imaging, coronagraphy, adaptive optics, active optics, exoplanets, binary stars, spatial light modulators, DAG telescope}

\section{INTRODUCTION}
\label{sec:intro}  

The Programmable Liquid-crystal Active Coronagraphic Imager for the DAG telescope (PLACID) will be the world's first ever  “adaptive stellar coronagraph” – “adaptive” describing the instrument’s unique properties of being adjustable to observing conditions in real-time. PLACID will be installed on a 4-m Ritchey-Chrétien telescope at the Eastern Anatolia Observatory (Do\u{g}u Anadolu G\"ozlemevi, short: DAG \cite{2022SPIE12182E..0XY}), incorporating a customized spatial light modulator (SLM) acting as a dynamically addressed focal-plane phase mask (FPM) coronagraph in the H – Ks bands. The PLACID project passed the Delivery Readiness Review (DRR) milestone in September 2023, and was delivered to Atatürk University Astrophysics Research and Application Center (ATASAM) facilities in March 2024. 
This new approach to high-contrast imaging is expected to have first light by late 2024 / early 2025, requiring early planning regarding its foreseen on-sky discovery space. Relying on an up-to-date exposure time calculator (ETC), first estimates of the science discovery space for PLACID are presented in this work. The considered targets are known directly imaged exoplanet and brown dwarf targets, considering parameters, such as raw lab contrast, contrast ratios, limiting magnitudes, coronagraphic inner working angle etc. The PLACID ETC and discovery space estimating tool are conceived to enable the Turkish and external astronomers to plan future observations.

\section{The PLACID instrument and DAG Telescope Observing Facilities}

\begin{figure} [ht]
   \begin{center}
   \begin{tabular}{c} 
   \includegraphics[height=8cm]{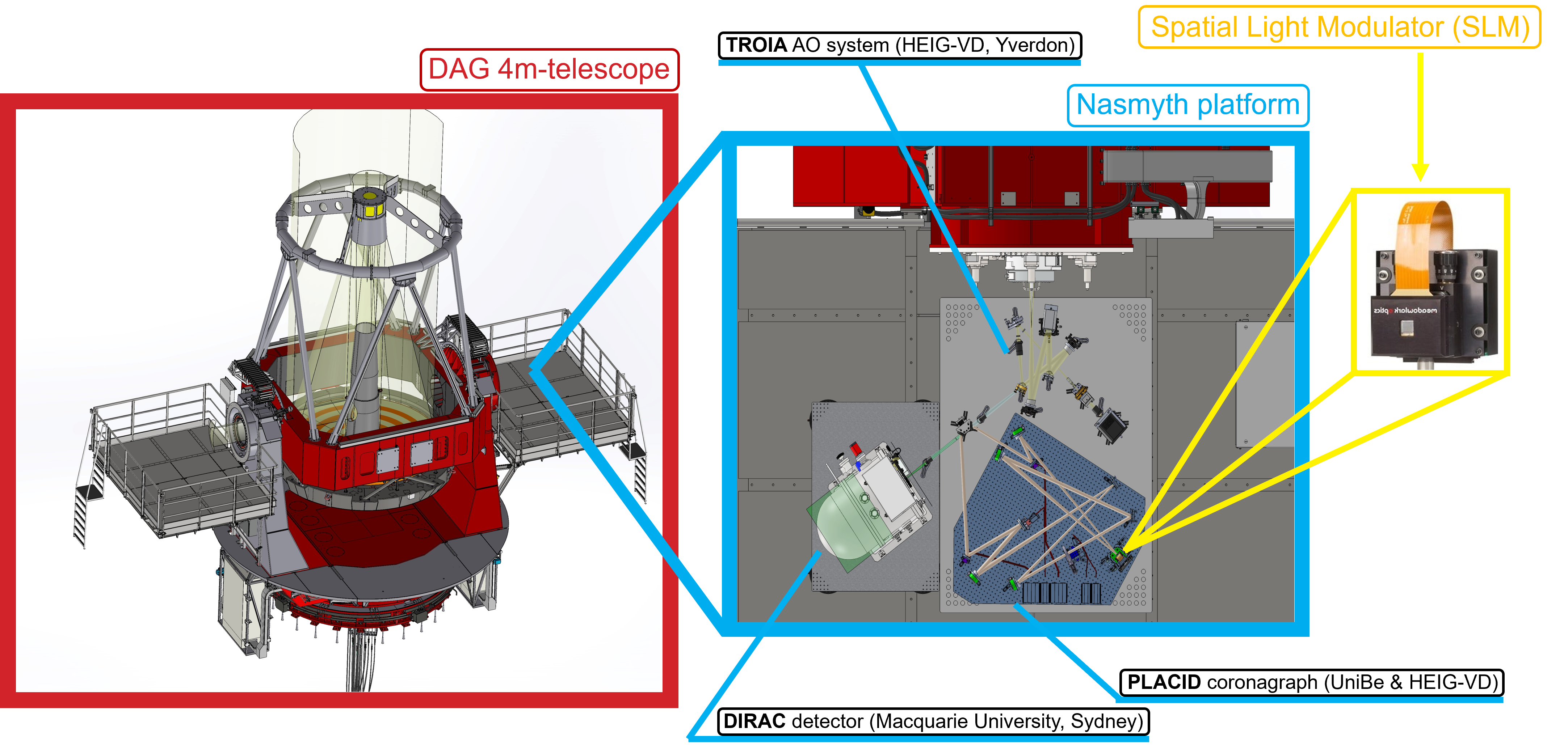}
	\end{tabular}
	\end{center}
   \caption[example] 
   { \label{fig:Nasmyth} 
The DAG telescope's Nasmyth platform, where PLACID will be installed (image credit: ATASAM, Meadowlark)}
   \end{figure}

\begin{figure} [ht]
   \begin{center}
   \begin{tabular}{c} 
   \includegraphics[height=6cm]{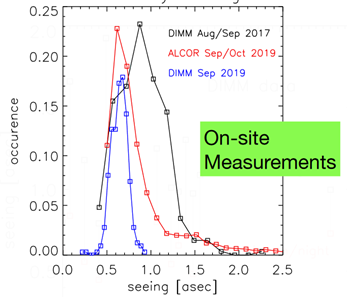}
	\end{tabular}
	\end{center}
   \caption[example] 
   { \label{fig:seeing} 
Expected seeing conditions at the DAG site}
   \end{figure}

The DAG observatory is Turkey's first large astronomical telescope, located at an altitude of 3170 m on Karakaya Ridge, close to the city of Erzurum. The dome harbours a 4m Ritchey-Chrétien telescope with an Alt-Az mount (see Table \ref{tab:DAG}). Two 4-by-3 m Nasmyth platforms are located on the telescope's sides, out of which one will provide seeing limited science in the visible, while the other provides adaptive optics in the NIR up to 2.4 $\mu$m. \cite{2021arXiv210203201K, 2022SPIE12185E..1WK} Furthermore, the instrument is equipped with an optical derotator (KORAY \cite{2020SPIE11445E..45K}) for compensating field rotation, as well as an active optics system on the primary mirror with 66 back and 12 lateral actuators \cite{2022SPIE12185E..1WK}. \\
The on-site observing conditions are expected to correspond to a median seeing of 0.9", with the best observed value of 0.3" (see Figure \ref{fig:seeing} for the measurements) and an average of 260/365 clear nights for a given year. \cite{2022SPIE12185E..1WK, 2021arXiv210203201K}
The adaptive optics system available on the NIR Nasymth is the TuRkish adaptive Optics system for Infrared Astronomy (TROIA), developed by HEIG-VD (University of Applied Sciences Western Switzerland, Yverdon-les-Bains, Switzerland). It is an extreme adaptive optics system (XAO) with a pyramid wavefront sensor (P-WFS) and an ALPAO deformable mirror (DM) with 468 actuators (see Figure \ref{fig:Nasmyth}). ALPAO DMs and their high stroke allow a correction of the turbulent tip-tilt as well as higher orders altogether, avoiding extra optics such as a tip/tilt (TT) mirror. The AO system uses a low-noise EMCCD camera and the expected TROIA residual wavefront errors are expected to be in the nm-range, dominated by astigmatism \cite{2022SPIE12185E..1WK}. The TROIA XAO system is expected to be installed at the site in the fall of 2024, more or less at the same time as the PLACID instrument.

\begin{table}[ht]
\caption{Basic data for the DAG telescope \cite{2022SPIE12185E..1WK, 2018SPIE10700E..2JY}} 
\label{tab:DAG}
\begin{center}       
\begin{tabular}{|l|l|} 
\hline
\rule[-1ex]{0pt}{3.5ex}  \textbf{Name} & \textbf{Do\u{g}u Anadolu Gözlemevi (DAG)}  \\
\hline
\rule[-1ex]{0pt}{3.5ex}  \textbf{Location} & Karakaya Ridge, Erzurum, Turkey   \\
\hline
\rule[-1ex]{0pt}{3.5ex}  \textbf{Latitude} & 39$^{\circ}$46'50.0" N   \\
\hline
\rule[-1ex]{0pt}{3.5ex}  \textbf{Longitude} & 41$^{\circ}$13'36.0" E   \\
\hline
\rule[-1ex]{0pt}{3.5ex}  \textbf{Altitude} & 3170 m   \\
\hline
\rule[-1ex]{0pt}{3.5ex}  \textbf{Seeing (lowest, median)} & 0.3" - 0.9" \\
\hline
\rule[-1ex]{0pt}{3.5ex}  \textbf{Primary mirror} & 4 m  \\
\hline
\rule[-1ex]{0pt}{3.5ex}  \textbf{Focal length} & 56 m  \\
\hline
\rule[-1ex]{0pt}{3.5ex}  \textbf{Mounting} & Altitude - Azimuth  \\
\hline
\rule[-1ex]{0pt}{3.5ex}  \textbf{Telescope build} & Ritchey - Chrétien  \\
\hline
\rule[-1ex]{0pt}{3.5ex}  \textbf{Declination limit} & $\geq$ -24$^{\circ}$  \\
\hline
\end{tabular}
\end{center}
\end{table}

The PLACID coronagraph follows downstream of TROIA on the optical table (Figure \ref{fig:Nasmyth}). TROIA uses visible light for wavefront sensing, while PLACID operates in the NIR, designed to work at H- (1.63 $\mu$m) to Ks-band (2.15 $\mu$m) with a customized compatible Lyquid Crystal on Silicon (LCOS) SLM (produced by Meadowlark). The pixelated SLM acts as a reflective, programmable focal plane mask (see yellow box in Figure \ref{fig:Nasmyth}), placed in such a way that a PSF spatial sampling of 10 SLM pixels per $\lambda/D$ at H-band are ensured. The device requires linearly polarized light in order to locally diffract away the undesirable stellar light, which will consequently be blocked by one of a variety of Lyot masks as part of a filter-wheel in the following pupil-plane. The instrument was jointly developed by the University of Bern and HEIG-VD.

The advantage of such an instrument is that it can be remotely configured on-demand to adapt to observing conditions, can conduct multiple star coronagraphy \cite{2018SPIE10706E..2NK} and correct self- calibrating non-common path aberrations (NCPAs) in real time. \\
Factoring in the resolution of the SLM (1920 x 1152 px), the field of view of the PLACID instrument will be set at 16" x 9.6" (see specifications in Table \ref{tab:PLACIDIRAC}). \\ 
Finally, the light path leads to the DAG InfRAred Camera (DIRAC), which is a Teledyne H1RG, hybrid CMOS detector with megapixel dimensions (18.3 mm x 18.3 mm and 1024 x 1024 px, see Table \ref{tab:PLACIDIRAC}), developed by Macquarie University, Sydney. DIRAC operates within a wavelength range of 0.9 - 2.4 $\mu$m, while also providing filter wheels that range from Y- to K-band. \cite{2022SPIE12184E..40Z} The minimum exposure time is expected to be of the order of 1s. \\
   
\begin{table}[ht]
\caption{Basic data for PLACID and DIRAC} 
\label{tab:PLACIDIRAC}
\begin{center}       
\begin{tabular}{|l|l|} 
\hline
\rule[-1ex]{0pt}{3.5ex}  \textbf{Name} & \textbf{PLACID}  \\
\hline
\rule[-1ex]{0pt}{3.5ex}  \textbf{Observing bands} & H-band: 1.63 $\mu$m (Ks-band: 2.15 $\mu$m)   \\
\hline
\rule[-1ex]{0pt}{3.5ex}  \textbf{SLM specification} & 1920 x 1152 px, 8 bits  \\
\hline
\rule[-1ex]{0pt}{3.5ex}  \textbf{FoV} & 16" x 9.6"  \\
\hline
\rule[-1ex]{0pt}{3.5ex}  \boldmath$\lambda/D$ \textbf{at H-band} & 85 mas \\
\hline
\rule[-1ex]{0pt}{3.5ex}  \textbf{Optical throughput} & $> 22$\%  \\
\hline
\hline
\rule[-1ex]{0pt}{3.5ex}  \textbf{Name} & \textbf{DIRAC} \\
\hline
\rule[-1ex]{0pt}{3.5ex}  \textbf{Detector} & Teledyne H1RG, 1024 x 1024 px \\
\hline
\rule[-1ex]{0pt}{3.5ex}  \textbf{Wavelength range} & 900 - 2400 nm \\
\hline
\rule[-1ex]{0pt}{3.5ex}  \textbf{Fill factor} & 100\% \\
\hline
\rule[-1ex]{0pt}{3.5ex}  \textbf{Minimum exposure time} & $\sim$ 1 s \\
\hline
\rule[-1ex]{0pt}{3.5ex}  \textbf{Read-out noise} & $\sim$ 18 e$^{-}$ \\
\hline
\rule[-1ex]{0pt}{3.5ex}  \textbf{Read-out gain} & 1.8 e$^{-}$/ADU \\
\hline
\end{tabular}
\end{center}
\end{table}

\section{The PLACID Discovery Space}

As PLACID is expected to be installed at the DAG by the end of this year, one of the first steps is to identify targets suitable for commissioning. \\
To this purpose, we compiled a list of feasible targets of previously directly imaged planetary mass companions (confirmed or candidates) and brown dwarfs, as well as circumstellar disks and binaries/multiple star systems. The latter systems will provide easier engineering targets, as the contrast between binaries is lower than between star and planetary mass companion/disk. \\ 
The location and physical parameters of the DAG, as well as the TROIA, PLACID and DIRAC instruments help us constrain the objects that can be observed.

   \begin{figure} [ht]
   \begin{center}
   \begin{tabular}{c} 
   \includegraphics[height=7cm]{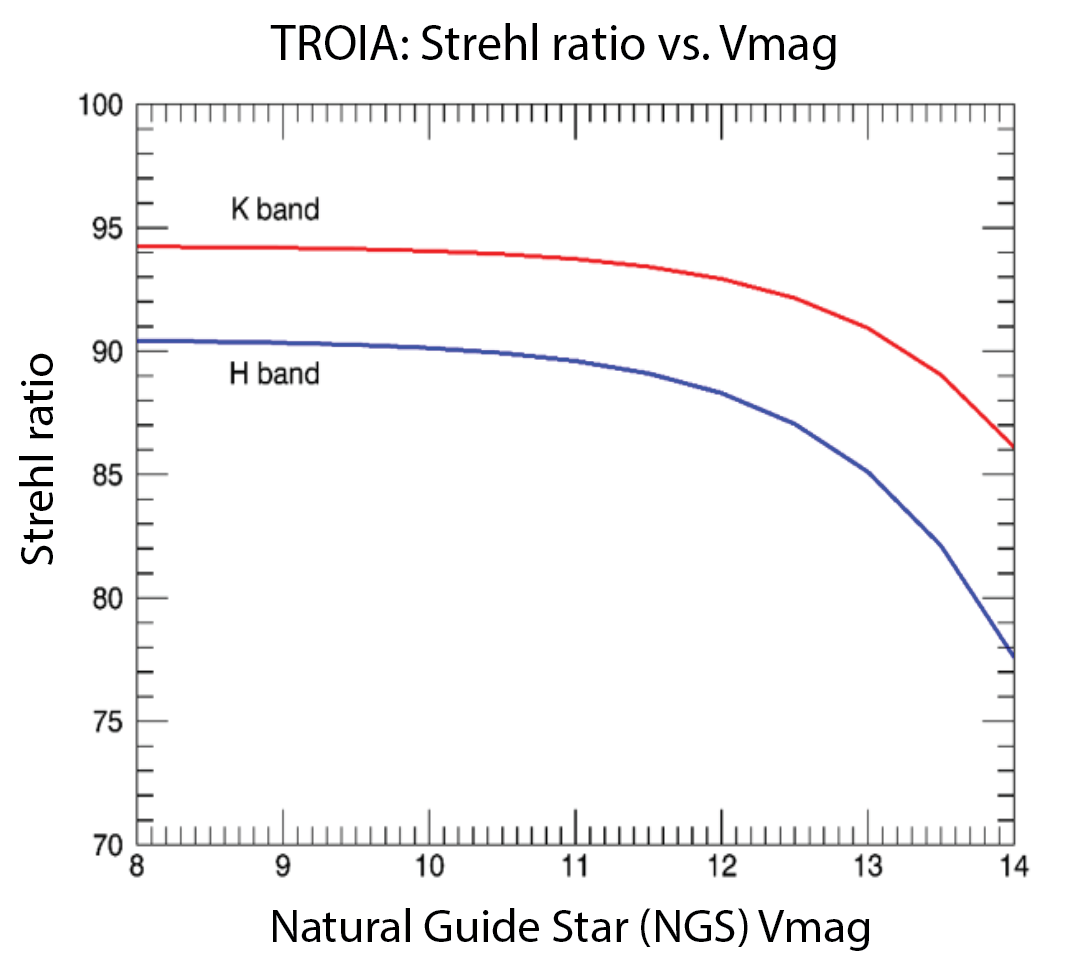}
	\end{tabular}
	\end{center}
   \caption[example] 
   { \label{fig:SRvsVmag} 
Expected sensitivity and correction performance using the TROIA AO system}
   \end{figure} 
   
First of all, the geographical location of the observatory, as well as airmass considerations constrain the accessibility of targets and lead to a declination limit of DEC $\geq$ $-24^{\circ}$. Furthermore, the host star visible magnitude required to close the loop with the TROIA XAO system is provided in Figure \ref{fig:SRvsVmag}. As previously mentioned, the AO system operates in the visible spectrum and is expected to provide an AO-corrected beam with a Strehl ratio between 85 - 95\% for the H- and K-band for magnitudes V $< 13$ mag, as indicated in Table \ref{tab:constraints}. \\
The third entry in the same table provides a constraint due to the field of view given by the instrumental configuration and the size of the SLM. Any object above 16" was excluded from the list, as it would not be possible to fit star and planet on the detector simultaneously.
   
\begin{table}[ht]
\caption{PLACID observational constraints} 
\label{tab:constraints}
\begin{center}       
\begin{tabular}{|l|l|} 
\hline
\rule[-1ex]{0pt}{3.5ex}  \textbf{Site} & DEC $\geq$ $-24^{\circ}$  \\
\hline
\rule[-1ex]{0pt}{3.5ex}  \textbf{TROIA AO guide star} & V $\leq$ 13 mag   \\
\hline
\rule[-1ex]{0pt}{3.5ex}  \textbf{On-sky FOV} & 16" x 9.6"  \\
\hline
\end{tabular}
\end{center}
\end{table}

\begin{figure} [!ht]
   \begin{center}
   \begin{tabular}{c} 
   \includegraphics[width=\textwidth]{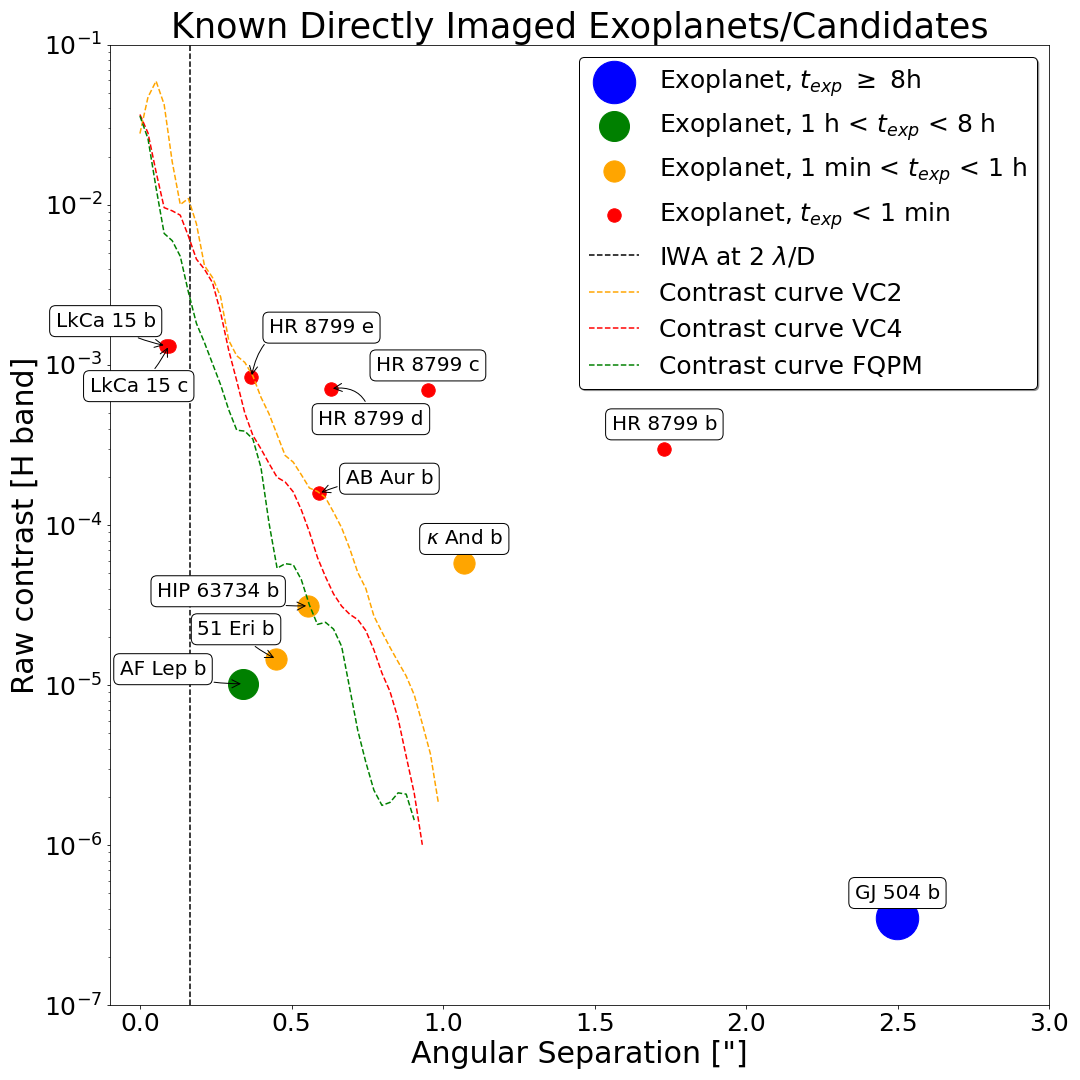}
	\end{tabular}
	\end{center}
   \caption[example] 
   { \label{fig:Exoplanet} 
Directly imaged exoplanets: angular separation vs. contrast including raw lab contrast curves \cite{2021arXiv210203201K} and expected exposure times for SNR = 5 (the coronagraph is not included in the simulation at this point)}
   \end{figure}

\begin{figure} [ht]
   \begin{center}
   \begin{tabular}{c} 
   \includegraphics[width=\textwidth]{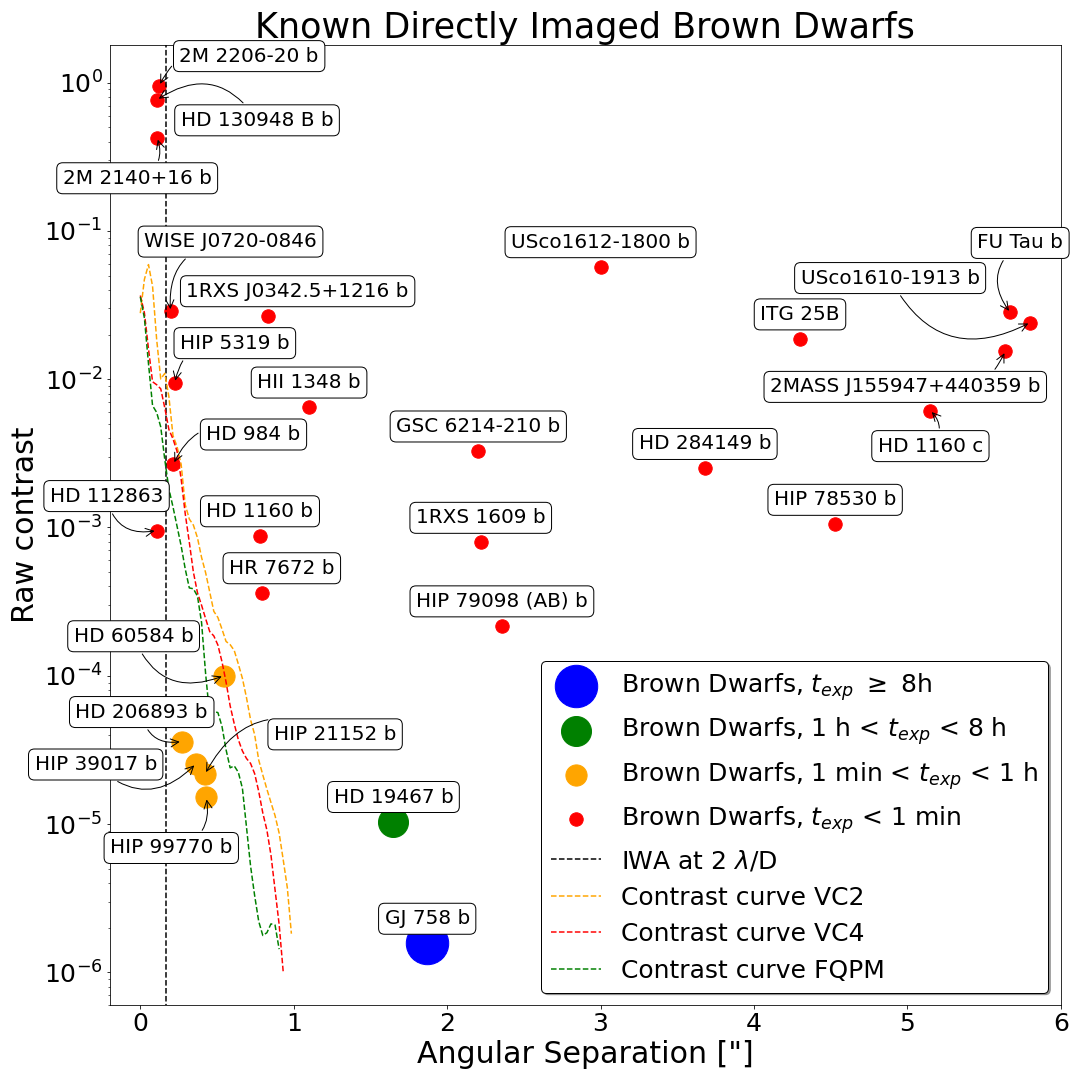}
	\end{tabular}
	\end{center}
   \caption[example] 
   { \label{fig:BrownDwarf} 
Directly imaged brown dwarfs: angular separation vs. contrast including raw lab contrast curves \cite{2021arXiv210203201K} and expected exposure times for SNR = 5 (the coronagraph is not included in the simulation at this point)}
   \end{figure}

The result of these observational constraints applied to known directly imaged exoplanets and brown dwarfs can be seen in Figures \ref{fig:Exoplanet} and \ref{fig:BrownDwarf}. These are plots showing angular separation vs. contrast ratio of star and companion. The dashed lines indicate the raw contrast values measured for PLACID in the lab \cite{2021arXiv210203201K}, with different colours for a vortex charge 2 focal plane mask (VC2), a vortex charge 4 FPM (VC4) and a four-quadrant focal plane mask (FQPM). The black dashed line indicates a conservative expected value for the Inner Working Angle (IWA).

Furthermore, a conservative exposure time estimate for each of the objects was calculated using e.g. expected transmission and Strehl ratio values as well as plate scale and read-out noise of the DIRAC detector. In this case, the expected exposure times are generated at given SNR = 5, as if the object was sitting atop the on-axis stellar PSF without any coronagraph, which is a very conservative scenario. This is however justified by the fact that raw PLACID contrast curves depicted on Figure \ref{fig:Exoplanet} are obtained in ideal laboratory conditions (SR $\sim$ 1), hence expected to worsen considerably once on-sky, although post-processing might then contribute positively. The estimation of the integration time at given SNR was calculated according to Nemati et al. 2020 \cite{2020JATIS...6c9002N}, without considering the significant effects of post-processing on the speckle noise.
   
As can be seen in Figure \ref{fig:Exoplanet}, the well-known systems of HR 8799 and $\kappa$ And are within reach of the PLACID coronagraph - HR 8799 e seems to be a borderline case, as it practically sits atop the VC2 lab contrast curve. The on-sky contrast curves will presumably degrade, however considering the post-processing would improve results roughly by a factor of 10, this could be considered a good estimate. \\
Similarly, Figure \ref{fig:BrownDwarf} provides a wide array of possibilities regarding the observability of brown dwarfs with PLACID, with most of the known directly imaged objects being within reach. In both graphs, a few objects can be found below the IWA-line, which will be interesting to probe, as they are good candidates to test PLACID's limits once on-sky measurements can be performed.


\section{CONCLUSION AND OUTLOOK}
\label{sec:conc}

The PLACID instrument was delivered to the ATASAM facilities in March 2024, while the DAG telescope and dome are expected to pass Final Acceptance over the summer of 2024. All instruments (including TROIA and DIRAC and the KORAY derotator \cite{2020SPIE11445E..45K}) are scheduled to be transported to the summit in the summer of 2024. As the AO Nasmyth platform will be equipped with a thermal enclosure this summer, the commissioning of PLACID and the other instruments on the platform is planned for fall 2024. In this time frame, the first light of PLACID can be foreseen for the end of 2024 or early 2025. \\
As we are moving closer to the installation of PLACID, the next steps are to refine the exposure time calculator with additional parameters, such as the case with a coronagraph, realistic speckle noise and the companion at an off-axis position. At the same time, the PLACID data reduction pipeline will be set up so that the first data can be processed by the end of the year. \\ 
For the instrument itself, the first step coming up is the commissioning of the instrument once it has been installed at the site. This will be followed by the first measurements of on-sky contrast by observing known objects and binaries.

   \begin{figure} [ht]
   \begin{center}
   \begin{tabular}{c} 
   \includegraphics[height=5cm]{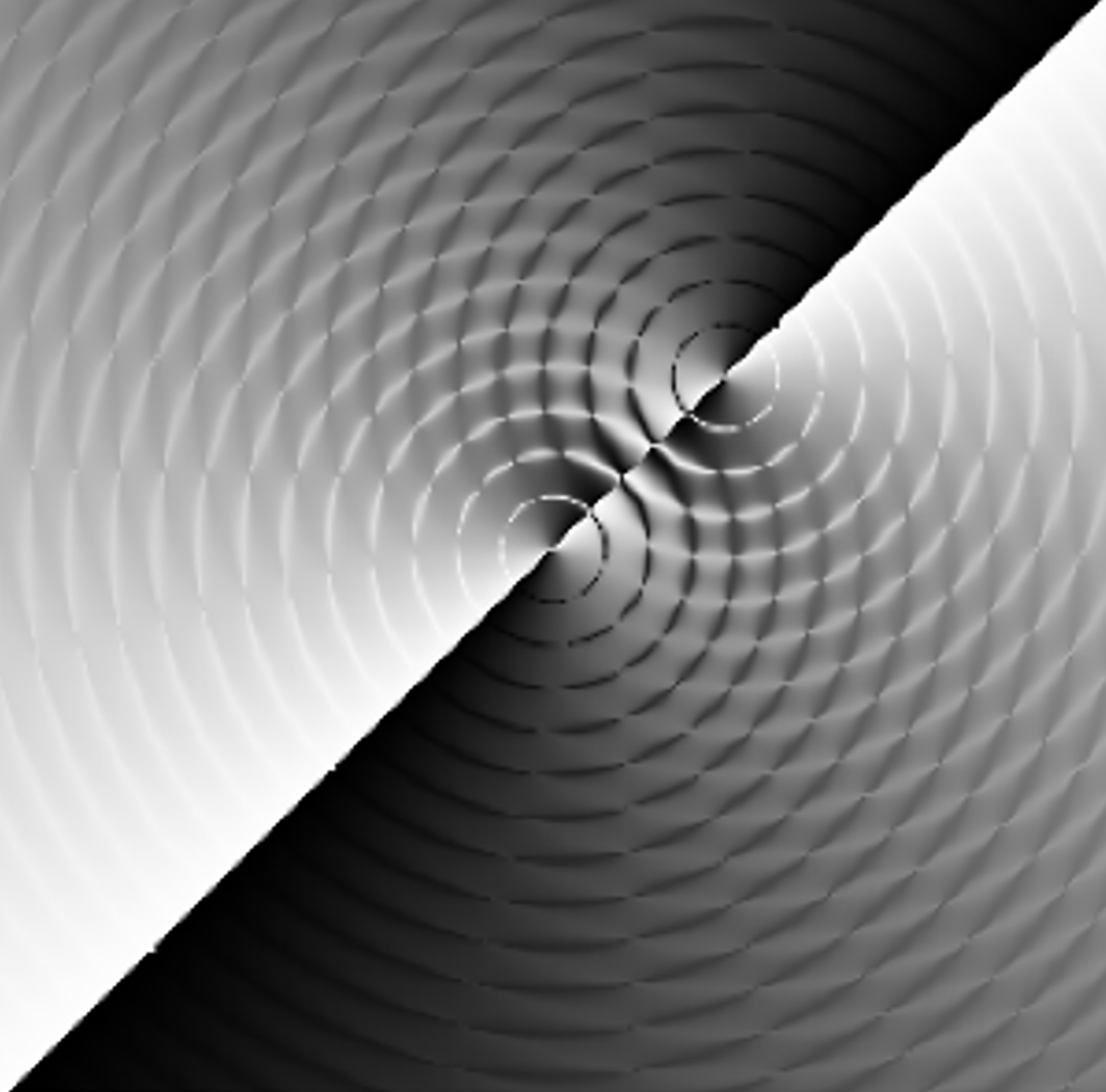}
	\end{tabular}
	\end{center}
   \caption[example] 
   { \label{fig:Binary} 
Vortex FPM (n=2) programmed for a binary star \cite{2018SPIE10706E..2NK}}
   \end{figure} 
   
Furthermore, the first Angular Differential Imaging (ADI) measurements can be implemented soon after that - a technique where the field derotator is tracking the pupil and the star can be subtracted from itself to leave behind the disk or companion. \\
Further steps following thereafter include the science that could be conducted with PLACID, such as detecting new companions, characterizing brown dwarfs (a type of object we do not know much about) and exoplanets, follow-ups of TESS and PLATO candidates for excluding false positives (eclipsing binaries, background stars, ...), characterizing circumstellar disks and lastly a niche science case with multiple star coronagraphy, where the SLM could be programmed to mask multiple stars and detect e.g. circumbinary disks or even planets (see Figure \ref{fig:Binary}). \\
This is of particular interest, as binary or multiple star systems represent about half of the stars in our galaxy. However, those are frequently excluded for high contrast imaging surveys, as they prove difficult to observe, particularly when using focal-plane coronagraphs and being confronted with companions with similar brightness and small angular separation. With the advent of the ELTs we will encounter many more close-in multiple star systems, as the angular resolution of the telescopes allow for such objects to be detected. \cite{2018SPIE10706E..2NK} It is therefore all the more important to investigate these systems at high contrast as they are so common. 

\acknowledgments 
 
The RACE-GO project has received funding from the Swiss State Secretariat for Education, Research and Innovation
(SERI), under the ERC replacement scheme following the discontinued participation of Switzerland to Horizon Europe.
Part of this work has been carried out within the framework of the National Centre of Competence in Research PlanetS
supported by the Swiss National Science Foundation under grants 51NF40 182901 and 51NF40 205606.  

\bibliography{proceedings} 
\bibliographystyle{spiebib} 

\end{document}